\documentclass[%
 aip,
 amsmath,amssymb,
 reprint,%
superscriptaddress
]{revtex4-1}
\setcitestyle{super} 
\usepackage{graphicx}
\usepackage{dcolumn}
\usepackage{bm}

\usepackage[utf8]{inputenc}
\usepackage[T1]{fontenc}
\usepackage{mathptmx}
\usepackage{etoolbox}
\usepackage{float}
\usepackage{upgreek}
\usepackage{xcolor}
\makeatletter
\def\@email#1#2{%
 \endgroup
 \patchcmd{\titleblock@produce}
  {\frontmatter@RRAPformat}
  {\frontmatter@RRAPformat{\produce@RRAP{*#1#2}}\frontmatter@RRAPformat}
  {}{}
}%
\makeatother
\usepackage{amsmath}
\newcommand{\NVminus}{$\mathrm{NV}^-$}
\newcommand{\NVzero}{$\mathrm{NV}^0$}
\newcommand{\gtwozero}{${g}^{(2)}(0)$}
\newcommand{\gtwot}{${g}^{(2)}{(\tau)}$}

\newcommand{\cminverse}{$\mathrm{cm}^{-1}$}
\newcommand{\highvacuum}{$\sim 5\times10^{-7}$~mbar}
\newcommand{\lowT}{6~K, $\sim 5\times10^{-7}$~mbar}
\newcommand{\Tone}{\textit{T}$_1$}
\newcommand{\Ttwo}{\textit{T}$_2$}
\newcommand{\alumina}{$\mathrm{Al}_2\mathrm{O}_3$}
\newcommand{\nanometer}{$\mathrm{nm}$}
\newcommand{\milliwatt}{$\mathrm{mW}$}

\begin{document}
\preprint{AIP/123-QED}

\title{Stabilisation of NV centres in diamond nanopillars at low temperature}
\author{Ravi Kumar$^{*}$}
\email{ucanrku@ucl.ac.uk}
\thanks{These authors contributed equally to this work.}
\affiliation{London Centre for Nanotechnology, University College London, London WC1H0AH, U.K.}

\author{Saksham Mahajan}
\thanks{These authors contributed equally to this work.}
\affiliation{Department of Electronic \& Electrical Engineering, University College London, London WC1E7JE, U.K.}

\author{Felix Donaldson}
\affiliation{London Centre for Nanotechnology, University College London, London WC1H0AH, U.K.}

\author{Leonardo Santoni}
\affiliation{Department of Chemistry, University College London, London WC1H 0AJ, U.K.}

\author{Aysha A. Riaz}
\affiliation{Department of Chemistry, University College London, London WC1H 0AJ, U.K.}

\author{Gediminas Seniutinas}
\affiliation{Qnami AG, CH-4132 Muttenz, Switzerland.}

\author{Felipe Favaro de Oliveira}
\affiliation{Qnami AG, CH-4132 Muttenz, Switzerland.}

\author{Anna Regoutz}
\affiliation{Department of Chemistry, University College London, London WC1H 0AJ, U.K.}
\affiliation{Department of Chemistry, University of Oxford, Inorganic Chemistry Laboratory, South Parks Road, Oxford OX1 3QR, U.K.}

\author{ Fabrizia Foglia}
\affiliation{Department of Chemistry, University College London, London WC1H 0AJ, U.K.}

\author{Siddharth Dhomkar}
\affiliation{London Centre for Nanotechnology, University College London, London WC1H0AH, U.K.}
\affiliation{Department of Physics, IIT Madras, Chennai 600036, India}
\affiliation{Center for Quantum Information, Communication and Computing, IIT Madras, Chennai 600036, India}

\author{John J.L.Morton}
\affiliation{London Centre for Nanotechnology, University College London, London WC1H0AH, U.K.}
\affiliation{Department of Electronic \& Electrical Engineering, University College London, London WC1E7JE, U.K.}

\date{\today}
\begin{abstract}
Degradation of near surface nitrogen vacancy (NV) centers in diamond under optical illumination has restricted their deployment in applications such as scanning NV magnetomety, particularly under harsh environment such as low temperatures and vacuum. Previously, alumina passivation of planar diamond samples has been shown to reduce the degradation of near surface ensemble NV centers in vacuum.
Here, we expand this study to incorporate photonic nanostructures by analyzing the single photon emission characteristics of NV centers embedded in an array of alumina-coated diamond nanopillars in vacuum (\highvacuum) and low temperature ($\sim$ 6~$\mathrm{K}$) environments under non-resonant (522~\nanometer) laser exposure. We find that, in contrast to the oxygen-terminated diamond nanopillars, NV centers in the alumina-coated nanopillars demonstrate negligible change in the single photon purity and brightness over the course of laser exposure in vacuum. At low temperature, NV centers under alumina termination demonstrate stable single photon emission, whereas under oxygen termination the single photon purity degrades under high intensity laser exposure. Alumina surface passivation is therefore shown as a viable path toward the realization of robust NV-diamond based nanoscale sensing under non-ambient atmospheric environments, including using diamond scanning probes.
\end{abstract}

\maketitle

\section{Introduction}

Quantum sensing at low temperature is a powerful approach for exploring nanoscale spin systems and emergent phases of condensed matter.\cite{casola2018probing,xu2023recent,degen2017quantum,rovny2024nanoscale,rondin2014magnetometry,lee2016ising,spaldin2021analogy,holenstein2016coexistence} To explore such fragile physical systems, a controlled sample environment under low temperatures is typically achieved under vacuum. In this context, the negatively charged nitrogen-vacancy (\NVminus) center in diamond has emerged as a leading platform, owing to its optically addressable spin states with long coherence times (\Ttwo~$\sim 2~\mathrm{ms}$) and ability to operate across a wide temperature range, from sub-Kelvin up to $\sim$ 600~K.\cite{gruber1997scanning,doherty2013nitrogen,balasubramanian2009ultralong,barry2020sensitivity,schirhagl2014nitrogen} Specifically, in scanning probe modalities, diamond probes incorporating single near-surface NV centers (typical depths $\sim$~10~\nanometer) enable nanoscale imaging with spatial resolution typically in the range of $\sim$ 20-50~\nanometer{} and magnetic field sensitivities of a few~$\upmu\text{T}\sqrt{\text{Hz}}$.\cite{xu2025minimizing,gross2017real} Using this approach, scanning NV magnetometry has enabled quantitative measurements of stray magnetic fields near material surfaces, including direct imaging of domain walls in antiferromagnets,\cite{appel2019nanomagnetism} and measurements of the London penetration depth in superconductors.\cite{thiel2016quantitative} Despite these successes, near-surface NV centers are increasingly recognized to suffer from performance degradation under non-ambient conditions. Under optical excitation, 
implantation-induced vacancy defects\cite{neethirajan2023controlled,pezzagna2010creation} and unstable surface functionalization\cite{kumar2024stability} lead to fluctuations in the NV charge state\cite{geng2023dopant,neethirajan2023controlled} and surface modifications.\cite{kumar2024stability,parthasarathy2024role} Such effects limit the applicability and performance of diamond scanning probes particularly in cryogenic and vacuum environments.\cite{kumar2024stability,parthasarathy2024role,neethirajan2023controlled,geng2023dopant}

Several strategies have been explored to mitigate the deterioration of near-surface NV-diamond sensors operating at low temperature and/or vacuum, including: low-power resonant laser excitation, the use of NV centers formed during diamond growth presenting better photostability,\cite{geng2023dopant} diamond surface adsorption of water,\cite{neethirajan2023controlled} and introducing oxygen in the sample chamber during optical measurements.\cite{parthasarathy2024role} However, these techniques have led to incomplete recovery of NV centers, as well as inducing potentially detrimental modifications in environmental conditions during sensing experiments. An alternative approach aimed to improve the photostability of near-surface NV centres, has been recently explored in ensembles of NV centers host in bulk diamond under vacuum,\cite{kumar2024stability} demonstrating diamond surface passivation using a thin layer ($\sim$ 2~\nanometer) of amorphous alumina (\alumina) to protect against degradation. This method has the advantages of a single fabrication step that produces, in a simple manner, a cap layer that shields the underlying NV centers from environmental influences while presenting negligible photoluminescence (PL) background. Importantly, the presence of such a cap layer has been demonstrated to negligibly influence the NV spin properties.\cite{liu2022surface,freire2023role} Nonetheless, the extension of the technique from ensembles in the bulk to single NV centres in nanopillars is non-trivial, given the difference in concentration of single substitutional nitrogen atoms (P1 centres), higher surface roughness in patterned diamond, \cite{xu2025minimizing} and difference in thermal dissipation, \cite{padgett2006thermal} as well as deposition-related constraints in regards to creating such cap layer on micro-structured surfaces with multiple structures presenting a plurality of surfaces and relative angles.

In this article we extend the application and explore the impact of alumina-based surface passivation on near surface single NV centers embedded diamond nanopillars,  exploring their behavior under optical illumination in vacuum (\highvacuum) and low temperature (\lowT). Under both vacuum and low temperature environments, the NV centers hosted in oxygen terminated nanopillars demonstrate strong degradation of their single photon purity upon laser exposure due to the appearance of fluorescent carbon-related species on the surface. In contrast, for NV centers hosted in alumina coated nanopillars, no measurable degradation of the single photon purity is observed for laser power below photoluminescence (PL) saturation under vacuum and up to maximum investigated power of 1.8~\milliwatt{} under low temperature. These results demonstrate a method which can be readily applied to achieve full protection of near surface NV centers through a surface passivation, suitable for robust operation in applications such as scanning NV magnetometry subject to cryogenic temperatures and vacuum. 

\section{Experimental methods}
\begin{figure}[t]
    \centering
    \includegraphics[width=0.49\textwidth]{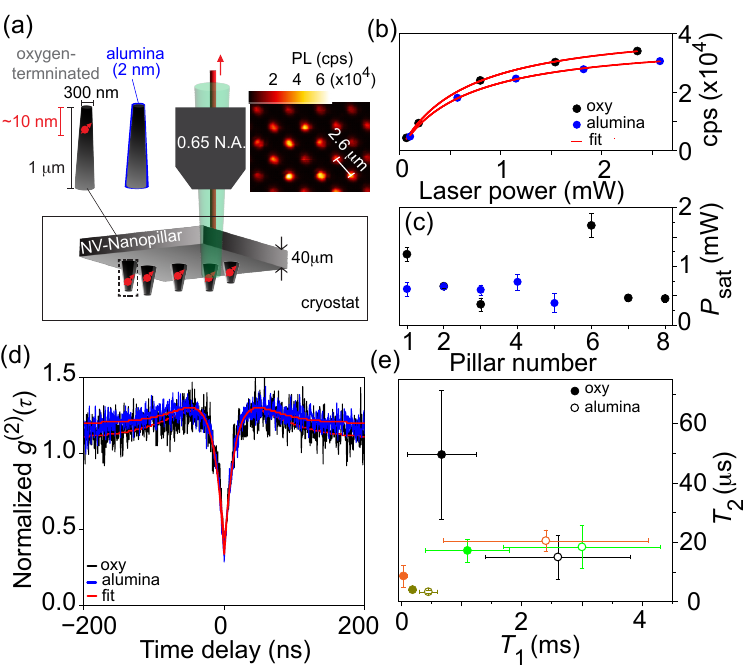}
    \caption{Optical characterisation of NVs in diamond nanopillars (NPs) under ambient conditions. (a) Sample and measurement configuration, showing NPs under two different surface terminations (oxygen-terminated and additional alumina coated). A confocal fluorescence image of a selected region of the sample (under oxygen termination) shows the NP array. (b) PL intensity (counts per second, cps) versus laser power for a given NP with fits used to extract saturation power ($P_{\rm sat}$), which is then plotted in (c) for selected NPs under different surface conditions. (d) Normalized \gtwot{} data for a selected NP under different surface terminations, with respective fit, used to confirm presence of a single NV emitter. (e) Longitudinal spin relaxation (\Tone) and spin coherence (Hahn echo \Ttwo) times for selected NPs (shown with different colours) under different surface terminations (indicated by open and closed circles).}
    \label{fig:figure 1}
\end{figure}

The sample used in the present study is a (100) electronic grade diamond ($3~\mathrm{mm}\times3~\mathrm{mm}\times40~\upmu\mathrm{m}$) which was implanted with 6~$\mathrm{keV}$ of $^{15}\text{N}^+$ ions followed by high temperature annealing (900~$^\circ$C, 2~hours) leading to the formation of near-surface NV centers at an average depth $\sim$ 10~\nanometer{}, estimated by stopping and range of ions in matter (SRIM)\cite{ziegler2010srim}. Thereafter, cylindrical nanopillars with 300~\nanometer{} (apex diameter) and $\times 1~\upmu\mathrm{m}$ (height) were etched out on the top of the implanted side, resulting into an average concentration of about one NV per nanopillar. Etched fiducial labels enabled identification of selected nanopillars so that the same nanopillar could be measured under different experimental conditions (Figure~S1). After the fabrication of nanopillars, the sample was acid cleaned and glued on the top of a silicon support frame (with the nanopillars facing the opposite side of the collection optics) using a temporary adhesive (crystal bond). The finally prepared sample was provided by Qnami AG\cite{QnamiAG} and is illustrated in Fig.~\ref{fig:figure 1}(a).

For oxygen functionalization of the diamond surface, the sample was treated with triacid mixture (70\% Perchloric acid, 95--97\% Sulphuric acid and 68\% nitric acid in a 1:1:1 v/v ratio) at 255~$^\circ\mathrm{C}$ for 2~hours.\cite{kumar2024stability} For alumina passivation, the triacid reflux above was followed by the deposition of a $\sim$ 2~\nanometer{} thick layer of \alumina{} through atomic layer deposition (ALD)  using a Savannah S200 system at 155 $^\circ\mathrm{C}$. To remove the alumina coating, the sample was soaked overnight in KOH (10$\%$ aqueous solution) at room temperature, rinsed with deionized (DI) water and subsequently treated with triacid mixture described above. To minimize the impact of laser-induced changes in the diamond surface and accommodate natural variations in NV properties (such as varying NV depth and surface morphology), we performed each new set of experiments on a previously unexposed sample region and (where possible) studied the identical nanopillar under different surface conditions. In some cases, the NV in an oxygen terminated nanopillar was found to have degraded, leading to the measurement of additional pillars. In this way, for each surface condition, at least five nanopillars were studied.

The optical measurements were performed using a Montana s100 cryostation integrated with a home-built confocal microscope. For efficient delivery of microwave field across a wide region and in order to address a large number of nanopillars, a keyhole resonator (with design adapted from elsewhere)\cite{sasaki2016broadband,donaldson2024nuclear} was fabricated with a resonance frequency around 2.87~$\mathrm{GHz}$. While the sample was placed inside the cryostat chamber, optical excitation was achieved using a continuous wave 522~\nanometer{} laser (LBX-522; Oxxius) through an air objective lens with 0.65~numerical aperture (N.A.) in the configuration shown in Fig.~\ref{fig:figure 1}(a). The fluorescence signal from the sample was guided through a dichroic mirror (550~\nanometer) followed by a 550~\nanometer{} long pass filter (LPF) to be coupled with either the photoluminscence spectrometer (SpectraPro HRS500, Princeton instruments) or single photon counting module (Excelitas Technologies). The NV spins were characterized in zero applied magnetic field: \Tone\ was measured by optical polarization of NVs followed by readout after some delay time, and \Ttwo\ was measured using a Hahn echo pulse sequence. For second order autocorrelation function measurements, the fluorescence signal was collected by two separate avalanche photo diodes (APDs) through a split core fiber, and the outputs were collected and analyzed with a Time Tagger~20 (Swabian instruments). Due to changes in the fluorescence background signal under optical illumination in vacuum and low temperature, background subtraction was not performed on the normalized \gtwot{} data, and instead accounted for such signal by fitting the acquired data using  $g^{({2})}_{\text{exp}}(\tau) = 1 - {\rho}^2 (1-g^{({2})}(\tau))$, where $g^{({2})}_{\text{exp}}(\tau)$ is experimental data, $g^{(2)}(\tau) = 1 - \beta e^{-\gamma_{1}\tau} + (\beta - 1)e^{-\gamma_{2}\tau}$, and $\rho=\tfrac{S}{S+B}$ ($S$ and $B$ represent the signal and background fluorescence counts, respectively). The parameters $\beta$, $\gamma_{1}$ and $\gamma_{2}$ are associated with decay rates in three level system and $\tau$ is delay time\cite{berthel2015photophysics}.

The properties of NVs under optical illumination were analyzed in ambient, vacuum (\highvacuum), and low temperature (\lowT) environments. To select an appropriate optical illumination intensity for laser exposure experiments, PL saturation measurements were performed on selected nanopillars under different surface conditions (Fig.~\ref{fig:figure 1}(b)) in ambient environment. The background subtracted PL saturation data were fitted using the relation $ I = \frac{P I_{\text{sat}}}{P + P_{\text{sat}}}$, where $I_{\text{sat}}$ and $P_{\text{sat}}$ denote the saturation PL intensity and laser power, respectively. 

Raman spectroscopy of the oxygen-terminated sample, after laser exposure under vacuum, was performed using a Bruker Raman microscope equipped with a 532~\nanometer{} laser (6.5 -- 12.5~\milliwatt{} of laser power was focused through a 0.5~N.A. air objective) under ambient atmosphere. Raman spectra of different nanopillars were recorded in the range 1000 -- 2000 \cminverse{}. Signal intensities (total area under the curve) of non-diamond carbon (1350 -- 1650~\cminverse) and diamond (1320 -- 1340~\cminverse) were compared to obtain the ratio $I_{\text{non-dia}}/I_{\text{dia}}$, indicating the relative abundance of both species.

\section{Results and Discussion}
Before analyzing the environmental impact of high-power optical illumination, we characterized a set of nanopillars under ambient conditions to probe the differences between the oxygen and alumina surface conditions (see Fig.~\ref{fig:figure 1}). Remarkably, at low optical power and ambient conditions, both surface conditions (oxygen or alumina) were found to have negligible impact on photon antibunching (\gtwot) characteristics.
The average values of $P_{\text{sat}}$ do not vary significantly for NV centers under the two different surface conditions. However, it is observed that the variance in $P_{\text{sat}}$ is lower for alumina-coated nanopillars, with ($P_{\text{sat}}(\text{avg,~oxy})=0.8 \pm 0.5~\mathrm{mW}$ and $P_{\text{sat}}(\text{avg,~alumina})=0.6 \pm 0.1~\mathrm{mW}$). 
In terms of the spin properties, no significant effect in \Ttwo{} was observed for NV centers under the two different surface conditions, in accordance with earlier reports\cite{liu2022surface}. In contrast, there was a consistent increase in the measured \Tone{} under alumina surface coating. Importantly, effects driven by changes in temperature during \Tone{} measurements under different surface conditions were excluded (see Figure~S2). This indicates that the alumina coating results in a passivation of the diamond surface and, consequently, in a reduction in high-frequency ($\mathrm{GHz}$) magnetic field noise arising from surface defects and/or adsorbates, as observed recently in hetero-functionalization of diamond and core-shell nanodiamonds.\cite{freire2023role,zvi2025engineering,barzegaramiriolya2025functionalized} Under high power laser exposure (1.8~\milliwatt) and in ambient environment, optically detected magnetic resonance (ODMR) spectroscopy (Figure~S2) confirmed that NV centers remained stable for both surface conditions.

\begin{figure}[t]
    \centering
    \includegraphics[width=0.45\textwidth]{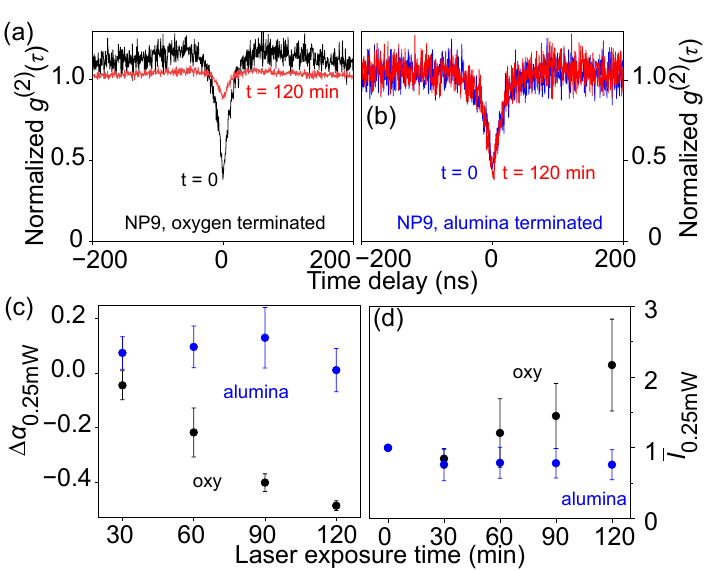}
    \caption{Effect of laser exposure under vacuum at a laser power of 0.25~\milliwatt. Normalized \gtwot{} data for Nanopillar-9 (NP9) at different laser exposure times under (a) oxygen and (b) alumina terminations. With reference to their values at $t=0$~ns, we show the time evolution under laser exposure of (c) the absolute change in single photon purity ($\Delta\alpha_{0.25~\mathrm{mW}}$) and (d) the relative change in PL count rate ($\overline{I}_{0.25~\mathrm{mW}}$). In (c) and (d), the data are average changes across five nanopillars, and error bars in (c) and (d) represent one standard deviation.}
    \label{fig:figure 2}

\end{figure}

In contrast to ambient conditions, the impact of laser exposure in a vacuum environment on the near-surface NV centers in nanopillars can be seen through the change in single photon purity ($\alpha = 1 - g^{(2)}{(0)}$), shown in Figs.~\ref{fig:figure 2}(a-c). 
Illustrative $g^{(2)}{(\tau)}$ traces obtained from one of the nanopillars (NP9) show a substantial reduction in single photon purity for the oxygen terminated sample after 120 min of laser exposure, while the same nanopillar, when alumina-coated, showed marginal changes over the same period.
Such behaviour is observed in multiple nanopillars, with average values obtained by studying five NPs (NP9--NP13, each measured first when alumina-coated and then when oxygen terminated).
The average change in single photon purity resulting from prolonged laser exposure is plotted (defined as $\Delta\alpha_{j}(t)=\frac{1}{5}\sum_{i=1}^{5}(\alpha_{ij}(t)-\alpha_{ij}(0))$, where $\textit{i}$ and $\textit{j}$ denote the pillar number and laser powers, respectively), confirming that the alumina coating is highly effective at preserving single photon purity at 0.25~mW. To further investigate the cause of the observed reduction in $\alpha$, the change in the fluorescence intensity of nanopillars is shown ($\textit{I}$) over the same period of laser exposure. Again, average values are studied with respect to the initial value: $\overline{I}_{j}(t)=\frac{1}{5}\sum_{i=1}^{5}\frac{I_{ij}(t)}{I_{ij}({0})}$. For the sake of simplicity in the analysis, relative rather than absolute changes are analyzed to account for effects such as the different collection efficiency for each nanopillar. The results show that the reduction in single photon purity seen for the oxygen-terminated nanopillars is accompanied by a considerable increase in background fluorescence, while there is no such background when the nanopillars are alumina coated.
In the Supplementary Information, further data for each individual nanopillars are shown in Figure~S4, and the effects of higher laser powers (up to 1~mW) are shown in Figure~S5. At higher laser powers (0.6~mW and 1~mW), the emergence of a background fluorescence (and reduction in single photon purity) can also be seen even for the alumina-coated nanopillars, albeit at a much slower rate than when oxygen-terminated. We note that such powers are at or above the saturation power for single NV center embedded in nanopillars.

\begin{figure}[t]
    \centering
    \includegraphics[width=1\linewidth]{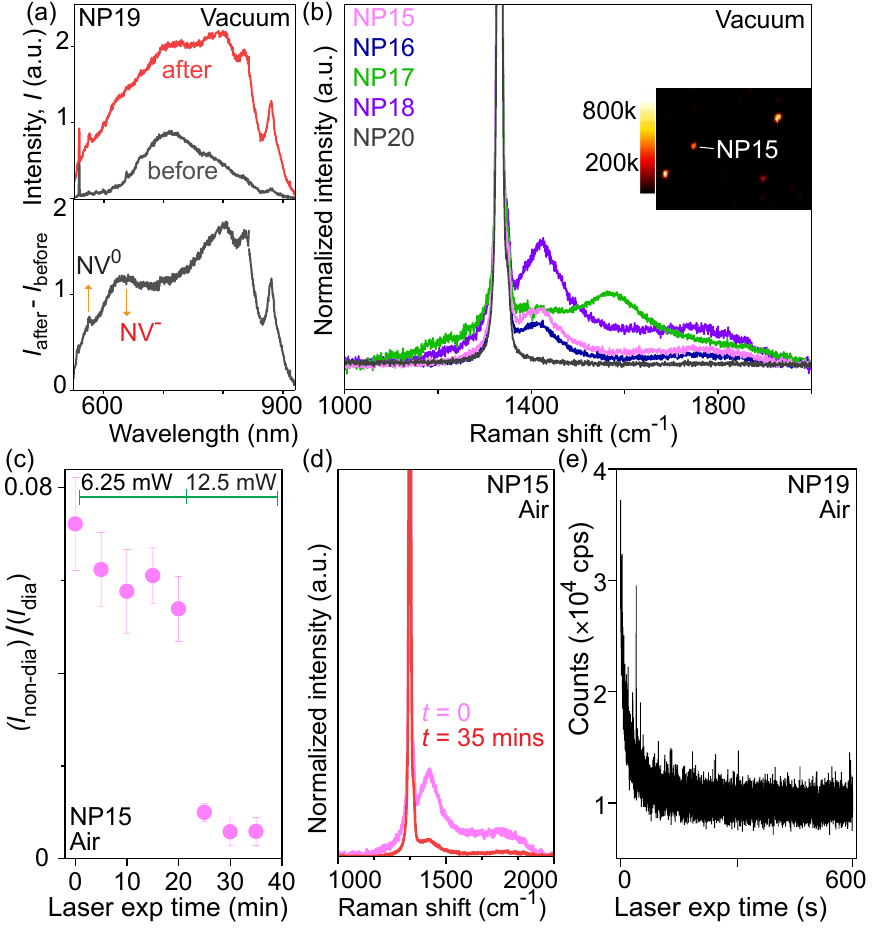}
    \caption{Surface reconstruction and NV charge state conversion in oxygen-terminated NPs in vacuum. (a) PL spectra of NP19 before and after laser exposure (0.25~\milliwatt{} for 180~min), with difference spectrum shown below. (b) Raman spectra of four NPs (NP15-18) following laser exposure and one unexposed pillar (NP20), the former indicating the appearance of non-diamond carbon. Inset shows a confocal PL image of NP15. (c) Time evolution of $I_{\text{non-dia}}/I_{\text{dia}}$ for NP15 in air under 532~\nanometer{} laser excitation at powers of 6.25 and 12.5~\milliwatt{}, respectively. (d) Raman spectra of NP15 before and after the 35 min of laser exposure defined in (c). (e) Photo-bleaching of vacuum exposed NP19 under optical illumination in air.}
    \label{fig:Figure 3}
\end{figure}

To understand the origin of laser-induced background signal observed for oxygen-terminated nanopillars and further probe the effects of the vacuum environment, we performed PL and Raman spectroscopy, as shown in Fig.~\ref{fig:Figure 3}, on a new (unexposed) set of nanopillars (NP15-20). The difference PL spectrum for NP19 --- obtained by subtracting the PL spectrum of unexposed NPs from that obtained followed laser exposure (0.25~\milliwatt, 180~min) under vacuum, confirms a broad increase in PL across the wavelength range of 550--900~\nanometer{} consistent with the background fluorescence discussed above. In addition, the difference spectrum shows a reduction of the intensity related to the \NVminus\ zero phonon line (ZPL) coupled with a marked increase in the \NVzero{} ZPL intensity, indicating a charge state transfer associated with a surface reconstruction and consistent with our observations on near-surface ensembles in vacuum environments\cite{kumar2024stability}.

We study changes in the diamond surface by examining the Raman spectra of four nanopillars following laser exposure, observing non-diamond carbon features that are absent in the spectra of an unexposed pillar (NP20). 
We rule out the possibility of a photo-induced carbon phase change ($sp^3\text{(diamond)} \rightarrow sp^2$) based on our previous measurements,\cite{kumar2024stability} and attribute this to the formation of fluorescent species (previously termed laser-induced carbon, LIC) arising from photo-polymerisation of organic molecule impurities.\cite{wagner2020laser,ling2009impact,parthasarathy2024role} 
The contribution of such non-diamond carbon (characterized by $I_{\text{non-dia}}/I_{\text{dia}}$), was subsequently reduced via in-situ optical illumination in air (Figures~\ref{fig:Figure 3}(c) and (d)) during Raman measurements, indicating its elimination in the presence of oxygen. The removal of the LIC is also supported by the observed reduction in fluorescence intensity from a nanopillar (NP19) in air, after it had been laser exposed in vacuum (see Figure~\ref{fig:Figure 3}(e)) as well as a reversal of the surface reconstruction, restoring the \NVminus\ charge state and high single photon purity (see Figure~S6). The observed stability of the NV optical properties within alumina coated nanopillars indicates that both of these detrimental effects are suppressed by the coating. First, the alumina coating inhibits the surface reconstruction under laser exposure in vacuum, as we previously observed in planar diamond surfaces\cite{kumar2024stability}. Second, there is an evident inhibition in the formation of LIC (for below-saturation illumination) which we attribute to the conformal ALD deposition of alumina, reducing the roughness of nanopillars and suppressing the adsorption of organic species. At laser powers higher than PL saturation, we observe an increase in the fluorescence background consistent with the degradation of the alumina coating and subsequent formation of LIC. 

\begin{figure}[t]
    \centering
    \includegraphics[width=1\linewidth]{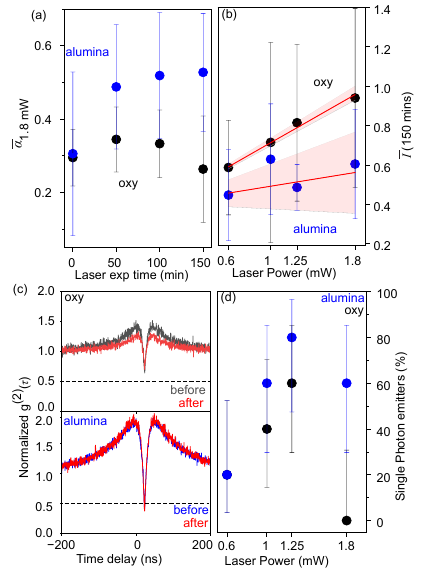}
    \caption{Low temperature (6~K) optical characterization. (a) Evolution of average single photon purity ($\overline{\alpha}$) under 1.8~\milliwatt{} laser illumination. (b) Effect of laser power on the average fluorescence intensity $\overline{I}$ measured after $150~\text{min}$ (normalised against the initial value) together with a linear fit (red curves). The red shaded regions represent the one-standard-deviation uncertainty in the fitted slopes. (c) Normalized \gtwot{} data of a nanopillar (NP21) under oxygen or alumina termination, before and after 1.8~\milliwatt{} laser exposure. The dotted lines denote the 0.5 threshold in \gtwozero{} for single photon emitters. (d) Fraction of single photon emitters (error bar represents 68.3~\% confidence interval estimated using Clopper-Pearson exact method due to small sample size) under different surface terminations after laser exposure for 150~min.}
    \label{fig:Figure 4}
\end{figure}

Having established the behaviour of NVs in nanopillars under vacuum at room temperature, further investigation is performed at cryogenic temperatures (6~K) under vacuum. Similarly, we focus on the evolution of absolute single photon purity and relative changes in fluorescence intensity of a new set of five nanopillars (NP21-25) under laser illumination to establish the impact of the different surface conditions. We define $\overline{\alpha}_{j}(t)=\frac{1}{5}\sum_{i=1}^{5}\alpha_{ij}(t)$ and $\overline{I}_{j}(t)=\frac{1}{5}\sum_{i=1}^{5}\frac{I_{ij}(t)}{I_{ij}(0)}$) to be the average values of single photon purity and normalized fluorescence intensity taken across five nanopillars under a given laser power $j$ and exposure time $t$.
The results are summarized in Fig.~\ref{fig:Figure 4}. At low temperature, optical illumination produced negligible change in $\alpha$ for oxygen terminated nanopillars (up to 1.8 mW), and a modest increase in $\alpha$ in the case of alumina coating. In contrast to the room temperature vacuum behaviour, where it was observed that the fluorescence background grew due to LIC formation, these changes were accompanied by a \emph{reduction} in fluorescence intensity of nanopillars (i.e.\ $\overline{I}<1$), likely due to surface desorption of fluorescent species rather than NV charge state conversion, which would have no impact on single photon purity.

Notably, there is a negative correlation between laser power and reduction in $\overline{I}$ particularly in the case of oxygen-terminated NPs (see Fig.~\ref{fig:Figure 4}(b)), attributed to the competing mechanism of LIC formation. The emergence of LIC at low temperature shows an approximately linear increase with laser power, similar to room temperature measurements, although the latter exhibits significantly higher LIC intensity (Figure~S11). We attribute this temperature dependence to a reduction in degassing from the diamond surface and in the rate of photo-polymerization of available organic species at low temperature. Under alumina surface passivation, LIC formation appeared either significantly slower or absent, for the exposure times and laser powers studied here. These observations indicate that LIC formation on diamond is common in vacuum environments, but its intensity strongly depends upon surface passivation, sample temperature and optical illumination conditions. The impact of emergent LIC on single photon purity is seen at 1.8~\milliwatt{}, where no single photon emitters (characterized by \gtwozero{} $<$~0.5) are observed among oxygen terminated nanopillars (Figs.~\ref{fig:Figure 4}(c) and (d)). In contrast, 3/5 of the investigated nanopillars under alumina termination were found to be single photon emitters under the same conditions. Raman spectroscopy performed on oxygen terminated nanopillars, exposed with 1.8~\milliwatt{} laser power under low temperature, reveal weak signatures of non-diamond carbon in some of the measured nanopillars, likely due to the small volume of LIC material (Figure~S11).

\section{Conclusion}
In conclusion, we find alumina coating by ALD provides an effective approach to mitigate the optically induced degradation of near surface NV centers in diamond scanning probes operated under low temperature and in vacuum. Our previous demonstration of enhanced charge-state stability of ensemble NV centers in planar diamond extends here to near surface single NV centers hosted in nanopillars used for scanning AFM probes. In air, the optical properties of NV centers remain insensitive to oxygen or alumina termination (modulo an increase in \Tone{} with alumina). Under vacuum at room temperature and below-saturation laser power (0.25~\milliwatt) the alumina coating is able to strongly suppress the formation of fluorescent organic matter which, for the oxygen-terminated samples, reduces the single photon purity. At higher laser powers, a reduction in single photon purity and increase in brightness is observed, albeit with a much slower rate than under oxygen termination. This suggests the gradual reconstruction of the alumina-terminated surface under high intensity laser illumination, which could form the basis of future studies. The behaviour is qualitatively different at low temperatures, where instead the single photon purity is found to increase for alumina-coated nanopillars under optical illumination, due to the more effective reduction in background fluorescence. As a result, a much larger fraction of single-photon emitters were observed for alumina-coated nanopillars compared to oxygen-terminated ones. In both room temperature and cryogenic regimes, these results support the use of amorphous alumina as a promising alternative to oxygen termination in order to achieve robust scanning NV-diamond probes for operation under vacuum conditions.

\section{Data availability}
The data supporting the findings of this study are available in the UCL Research Data Repository, DOI: 10.5522/04/32113225.

\section{Acknowledgments}
This work was supported by the Engineering and Physical Sciences Research Council (EPSRC) through the Hub in Quantum Computing and Simulation (Grant No. EP/T001062/1), EP/T517793/1 and Q-BIOMED Hub (Grant No. EP/Z533191/1). FF acknowledges EPSRC for funding (EP/V057863/1). The authors thank Antilen Jacob (London Centre for Nanotechnology, UCL, London WC1H0AH, U.K.) and LCN cleanroom team (London Centre for Nanotechnology, UCL, London WC1H0AH, U.K.) for their support during experiments. The authors thank Andrea Sella (Department of Chemistry, UCL, London WC1H 0AJ, U.K.) for helpful suggestions during chemical treatment of diamond.

\section{References}
\bibliographystyle{aipnum4-1} 
\bibliography{references}     

\end{document}


\preprint{AIP/123-QED}
\renewcommand{\thefigure}{S\arabic{figure}}
\renewcommand{\thetable}{S\arabic{table}}
\setcounter{figure}{0}
\setcounter{table}{0}

\title{Supporting Information}
\maketitle

\onecolumngrid
\section{Selection of nanopillars for optical characterization}
\begin{figure}[H]
    \centering
    \includegraphics[width=0.5\textwidth]{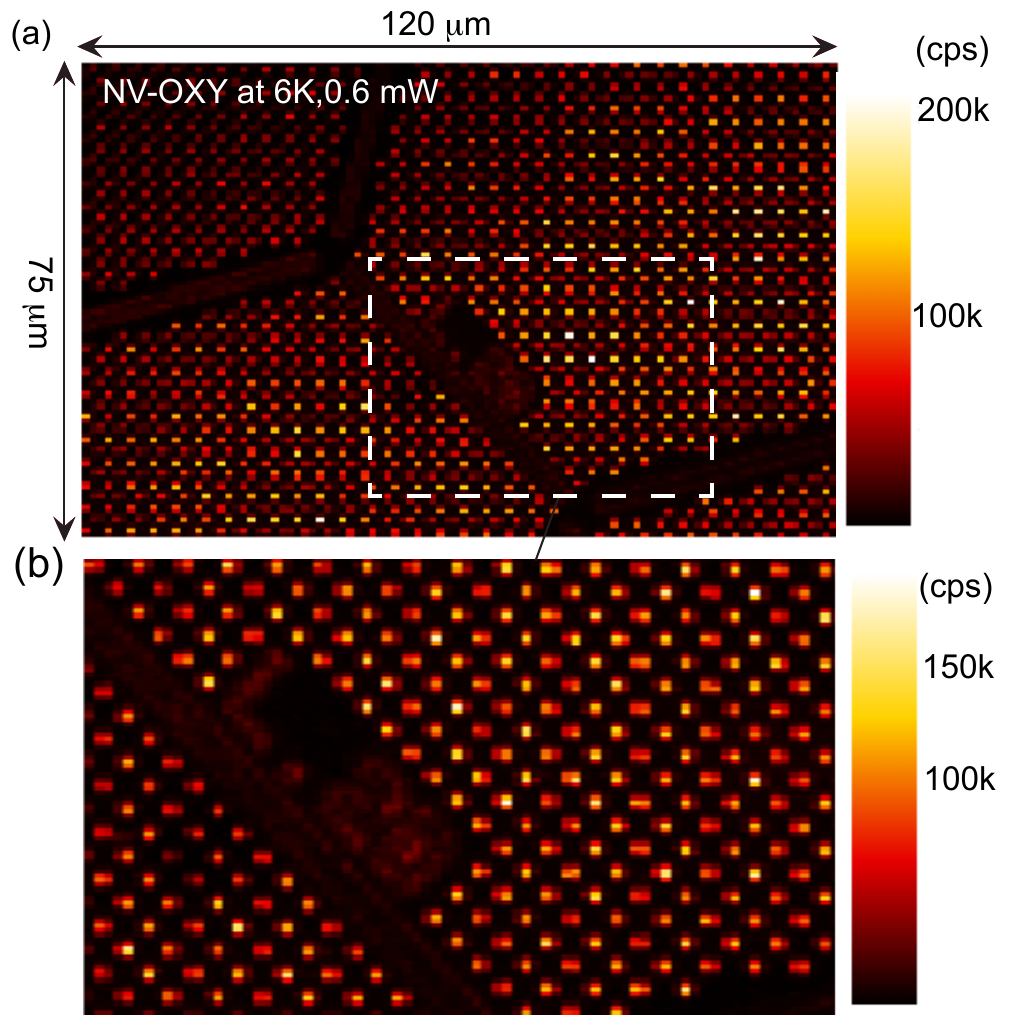}
    \caption{(a) Scanning confocal image of a selected region of the sample (under oxygen termination) (with field number L.20) under low temperature. The region around the field label (within dotted region) has been shown in (b) where each bright spot represents individual nanopillar with average concentration of $\sim$~1 NV/pillar at a nominal depth of about $\sim$ 10~\nanometer.}
    \label{fig:S1}
\end{figure}
\newpage
\section{Optical and spin characterization under ambient conditions}
\begin{figure}[H]
    \centering
    \includegraphics[width=0.65\textwidth]{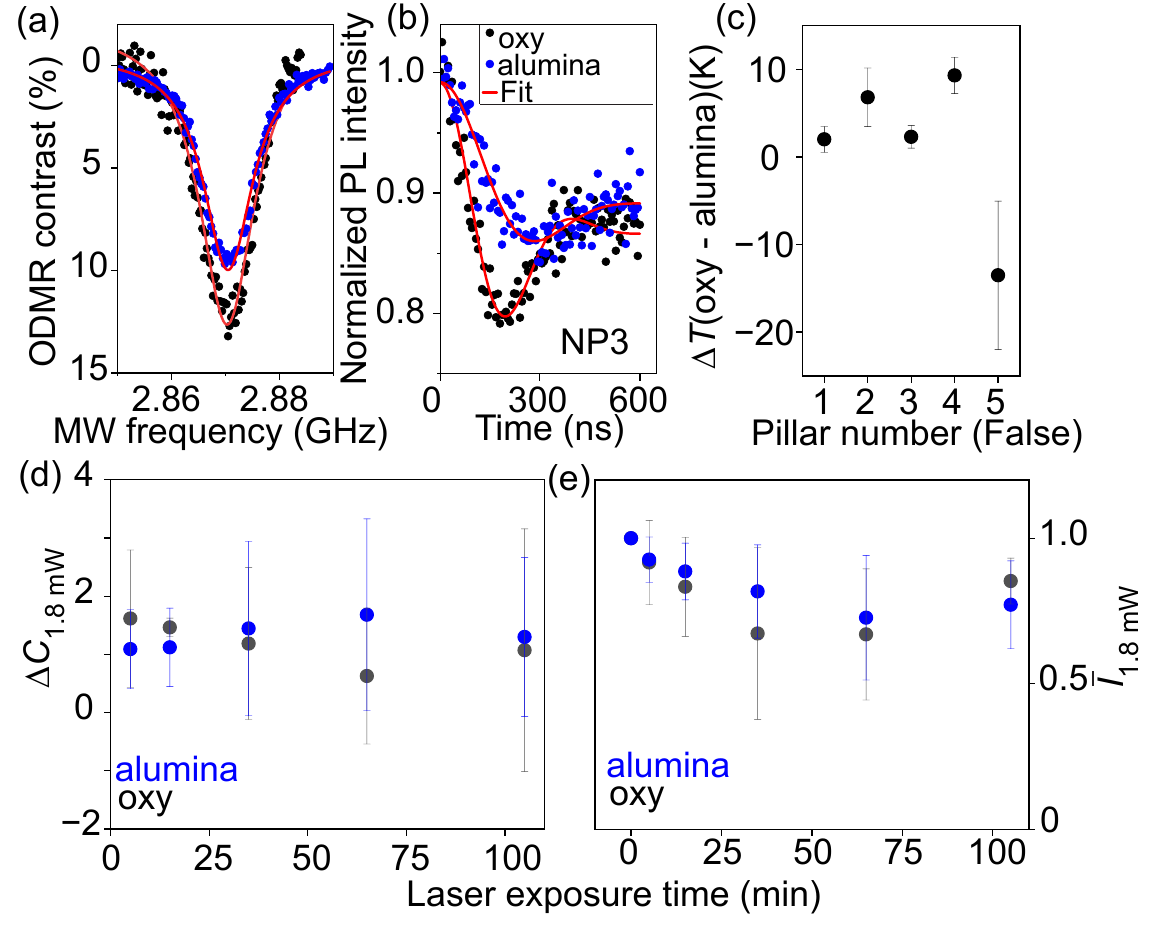}
    \caption{Optical and spin characterization under ambient conditions. (a) Continuous wave (CW) ODMR spectra of a selected nanopillar under oxygen and alumina passivation, respectively, along with fitted curve (Lorentzian line shape). The difference in ODMR contrast between oxygen and alumina passivation conditions is caused by the relative positioning of the sample on the resonator (leading to a change in \textit{B}$_1$ field intensity). (b) Rabi measurement  of the same nanopillar under different surface terminations along with the fitted curve ($I(t) = A \cos(2\pi f_\text{rabi}t)\, e^{-\left(\frac{t}{T_2^*}\right)^2} + C$ where $A$ is amplitude, $f_\text{rabi}$ is Rabi frequency, $T_{2}^{*}$ is spin dephasing time and $C$ is constant offset). (c) Relative temperature difference between oxygen and alumina surface passivation conditions during $T_\text{1}$ measurements. The change in temperature was calculated by first estimating the difference in zero field splitting (D) position ($\Delta D= {D}_{\text{OXY}} - {D}_{\text{Alumina}}$) and then correlating it with the temperature using the formula $\frac{dD}{dT}=-74.2~\mathrm{\frac{kHz}{K}}$.\cite{acosta2010temperature} (d) Time evolution of change in CW ODMR contrast of NPs under 1.8 \milliwatt{} laser exposure. Here, $\Delta C_{j}(t)=\frac{1}{5}\sum_{i=1}^{5}(C_{ij}(t)-C_{ij}(0))$ reflects the average change in contrast over the course of laser exposure. Whereas, $i$ and $j$ denote pillar number and laser power respectively. (e) Average change in brightness w.r.t. initial value, defined by $\overline{I}_{j}(t)=\frac{1}{5}\sum_{i=1}^{5}\frac{I_{ij}(t)}{I_{ij}(0)}$. Here, $I_{ij}(t)$ denotes the count rate of the $i^{{th}}$ pillar after $t$ min of $j$ \milliwatt{} of laser power.}
    \label{fig:S2}
\end{figure}

\begin{figure}[H]
    \centering
    \includegraphics[width=0.65\textwidth]{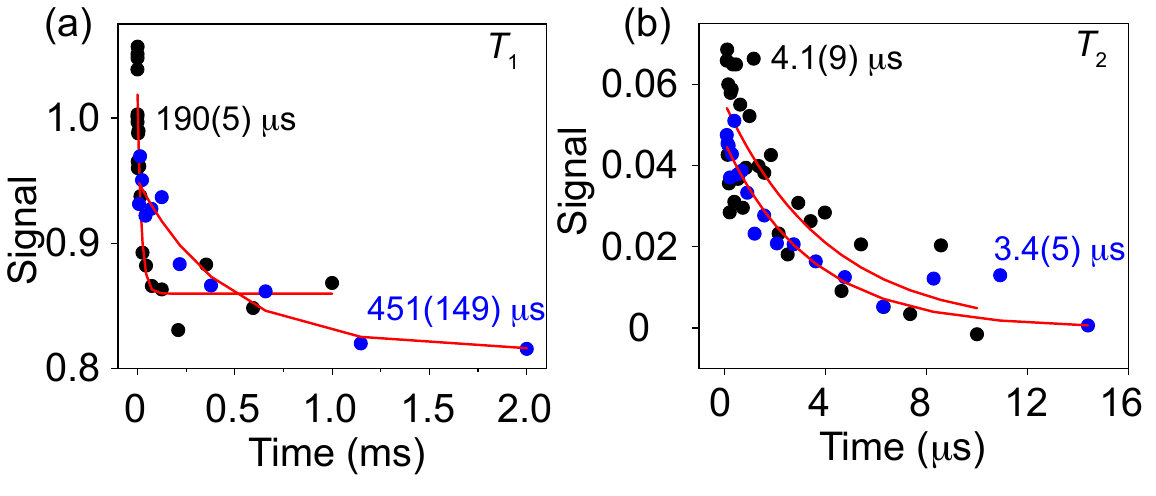}
    \caption{(a) longitudinal (\Tone) and (b) transverse (\Ttwo) relaxation time measurements for a selected nanopillar under different surface termination conditions in ambient environment.}
    \label{fig:S3}
\end{figure}
\newpage
\FloatBarrier
\section{Optical characterization under high vacuum}
\begin{figure}[H]
        \centering
        \includegraphics[width=1\textwidth]{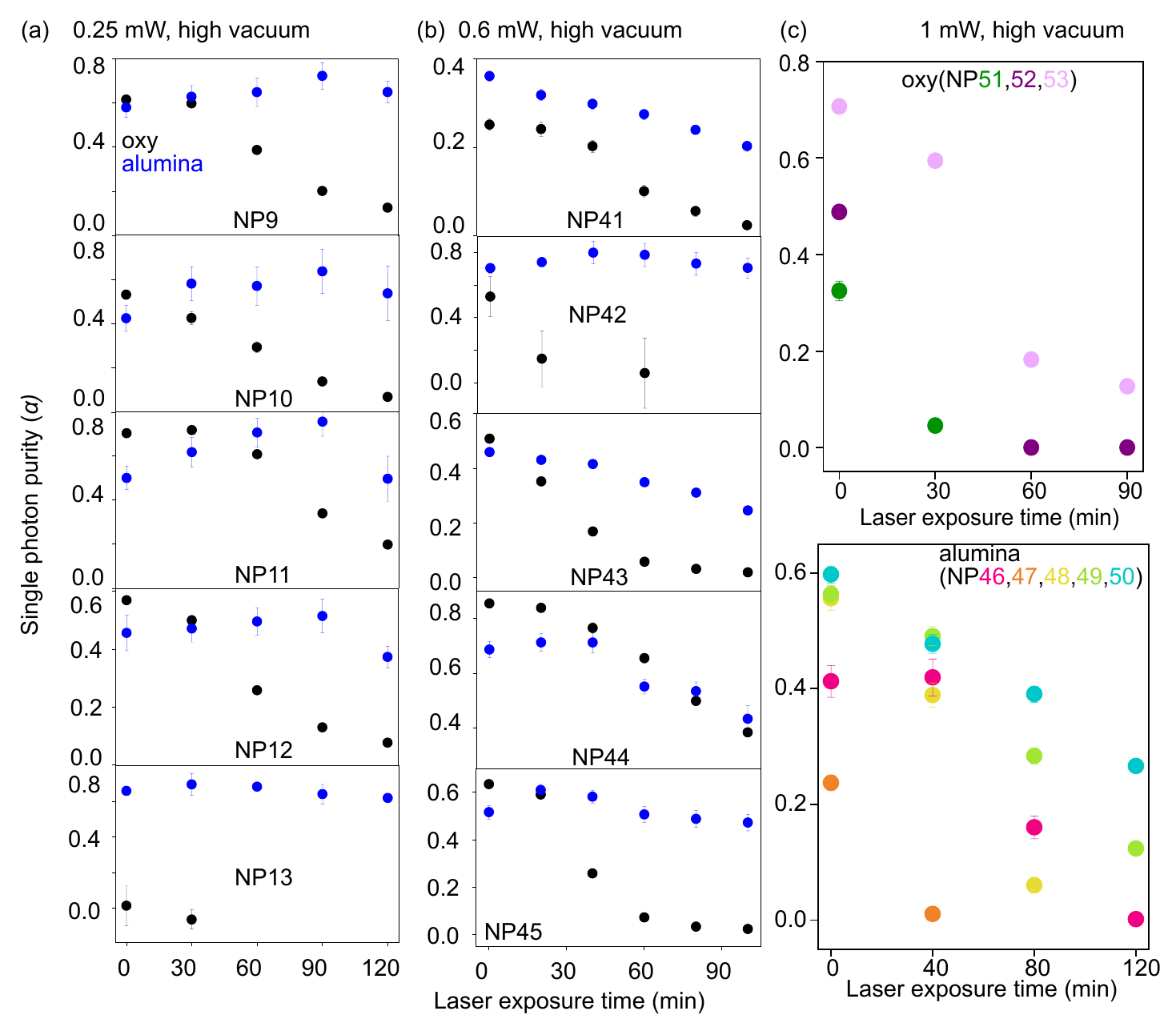}
        \caption{Comparison of change in single photon purity ($\alpha$) among selected nanopillars under different surface terminations in high vacuum environment. For each laser power, a separate set of about five previously unexposed NPs was selected and measured under different surface termination conditions. The change in $\alpha$ was observed at (a) 0.25~\milliwatt{}, (b) 0.6~\milliwatt{}, and (c) 1~\milliwatt{} laser power, respectively. At 1~\milliwatt{} laser power, a different set of NPs was measured under oxygen and alumina terminations}
        \label{fig:S4}
    \end{figure}
\begin{figure}[!t]
    \centering
    \includegraphics[width=0.65\linewidth]{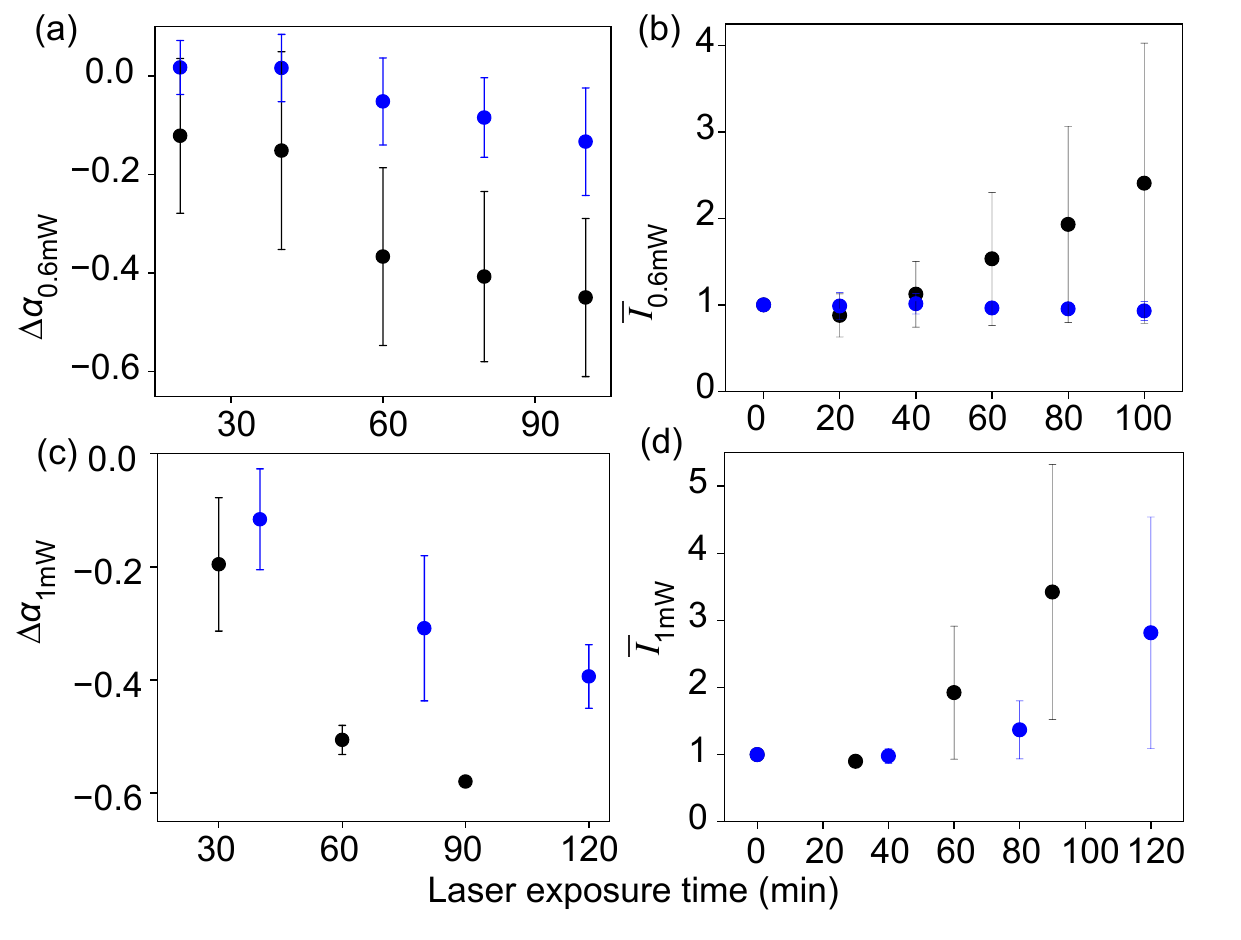}
    \caption{(a) Change in single photon purity under optical illumination in high vacuum at (a) 0.6~\milliwatt{} and (c) 1~\milliwatt{}, respectively. (b) and (d) represent corresponding change in brightness at laser powers of 0.6 and 1~\milliwatt{}, respectively.}
    \label{fig:S5}
\end{figure}
\begin{figure}[!t]
    \centering
    \includegraphics[width=0.65\linewidth]{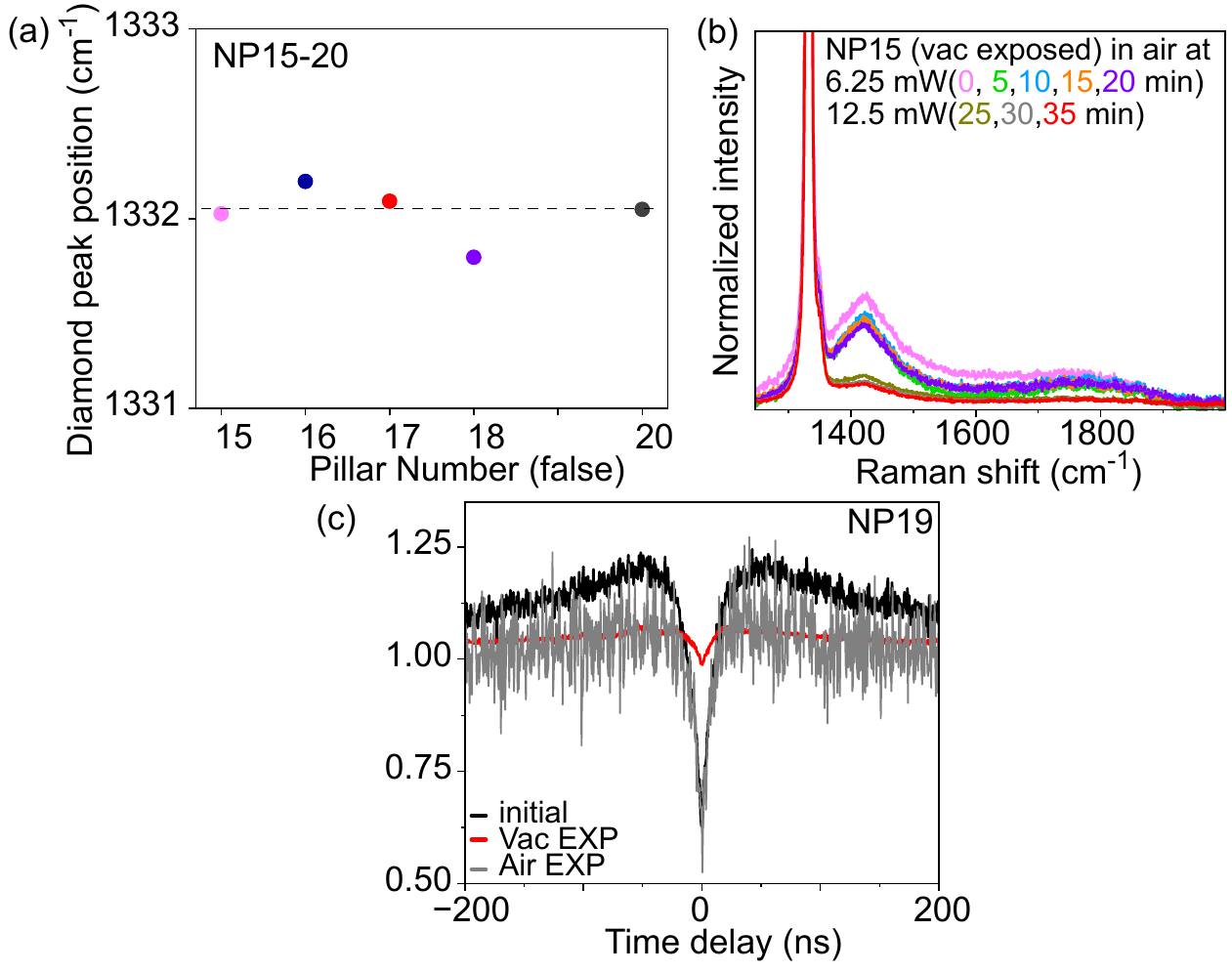}
    \caption{(a) Diamond peak position for different NPs laser exposed under high vacuum. The diamond peak position was measured by fitting the Raman spectra of different NPs using a Lorentzian line shape. (b) Raman spectra of NP15 (after laser exposure in vacuum) in ambient environment under different laser powers and exposure time. (c) The \gtwot{} of NP19 under high vacuum followed by laser exposure in vacuum and air, respectively. The degraded \gtwot{} features due to laser exposure in vacuum were recovered under ambient environment indicating reversible and atmosphere dependent characteristics of organic contamination (LIC).}
    \label{fig:S6}
\end{figure}

\newpage
\FloatBarrier
\section{Optical characterization at low temperature}
\begin{figure}[H]
    \centering
    \includegraphics[width=0.5\linewidth]{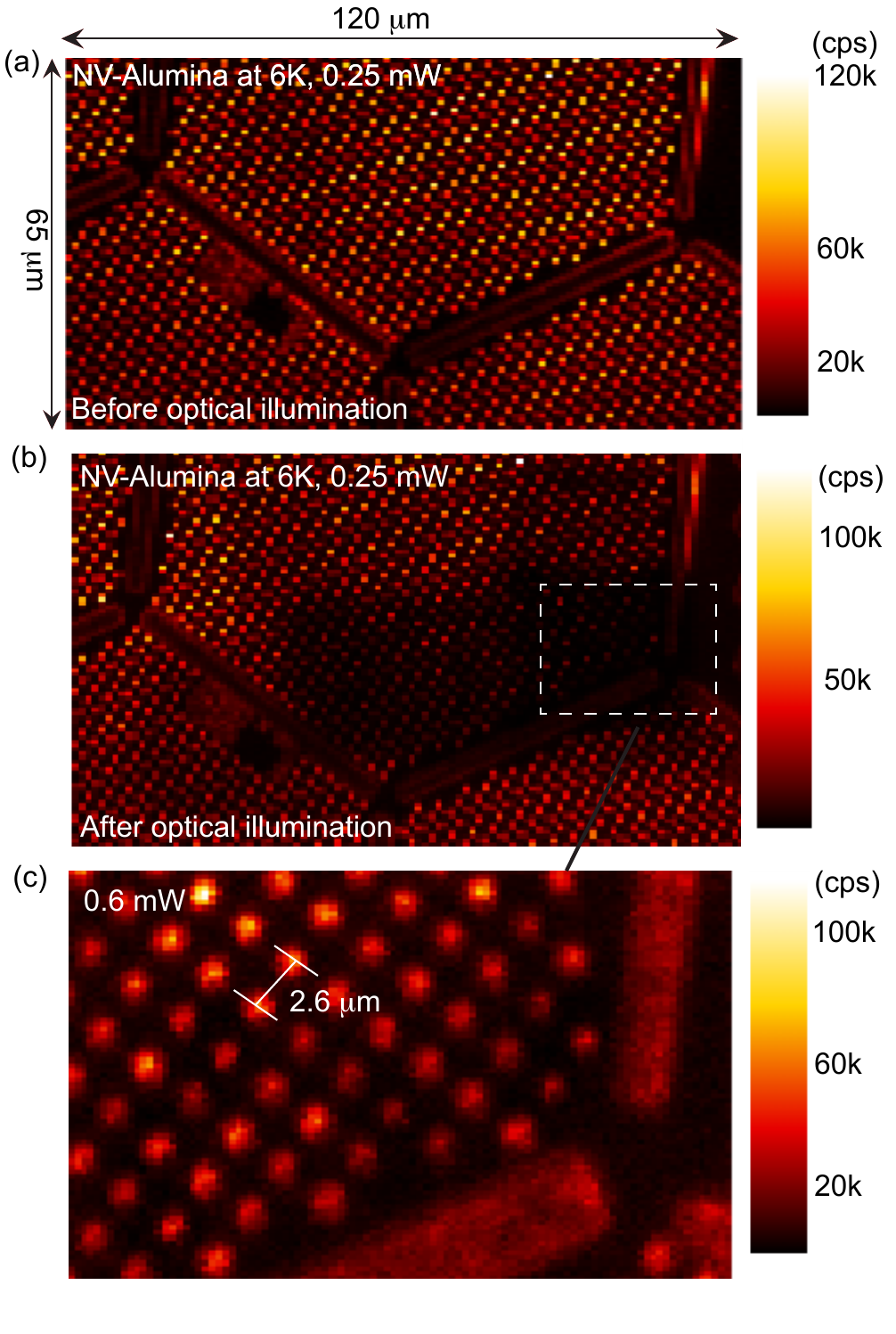}
    \caption{Optical image of a NV-Alumina sample region under low temperature (a) before and (b) after optical illumination measurements at laser power of 1.8~\milliwatt{}. The imaging was performed at a laser power of 0.25~\milliwatt{}. The region under the white dotted square was imaged at 0.6~\milliwatt{} and is shown in (c).}
    \label{fig:S7}
\end{figure}
\begin{figure}[!t]
    \centering
    \includegraphics[width=1\linewidth]{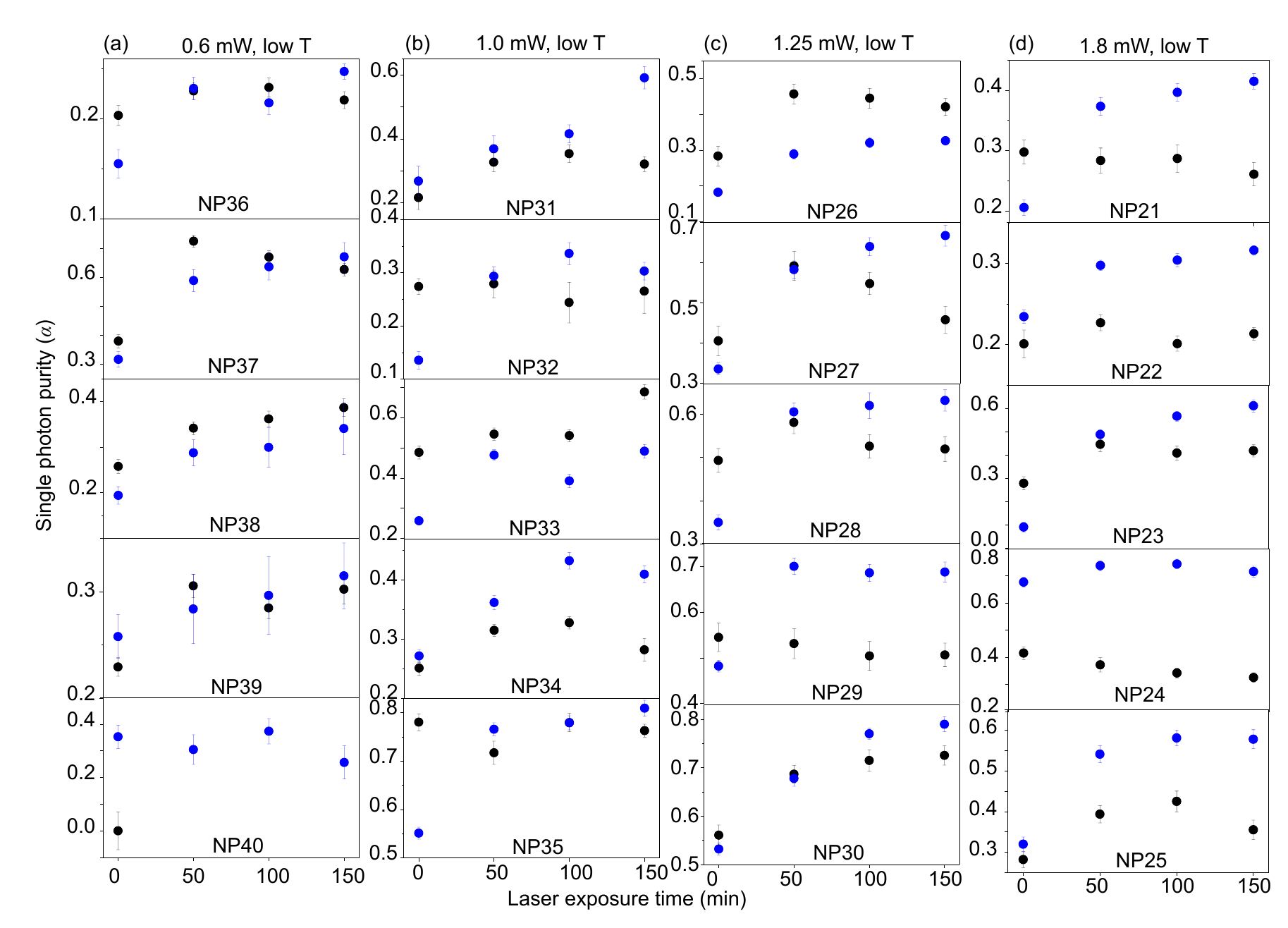}
    \caption{Change in single photon purity ($\alpha$) under laser exposure in a low temperature environment. The \gtwot{} was measured at the laser powers of (a) 0.6~\milliwatt, (b) 1~\milliwatt{}, (c) 1.25~\milliwatt{}, and (d) 1.8~\milliwatt, respectively.}
    \label{fig:S8}
\end{figure}
\begin{figure}[!t]
    \centering
    \includegraphics[width=1\linewidth]{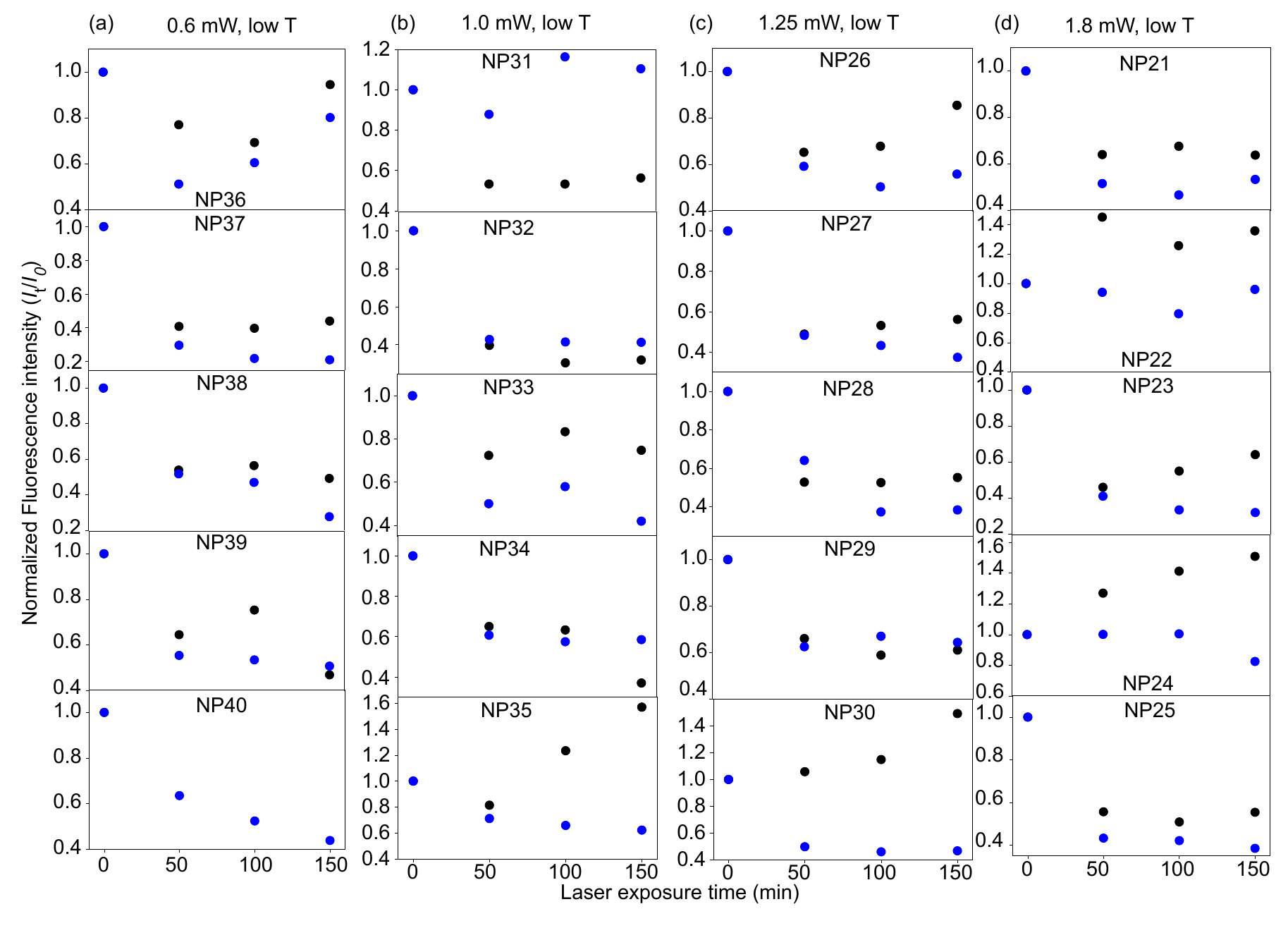}
    \caption{Evolution of normalized fluorescence intensity ($\frac{I_t}{I_0}$) under laser exposure in a low temperature environment. The fluorescence intensity was measured at the laser powers of (a) 0.6~\milliwatt, (b) 1~\milliwatt{}, (c) 1.25~\milliwatt{}, and (d) 1.8~\milliwatt, respectively.}
    \label{fig:S9}
\end{figure}
\begin{figure}[!t]
    \centering
    \includegraphics[width=0.75\linewidth]{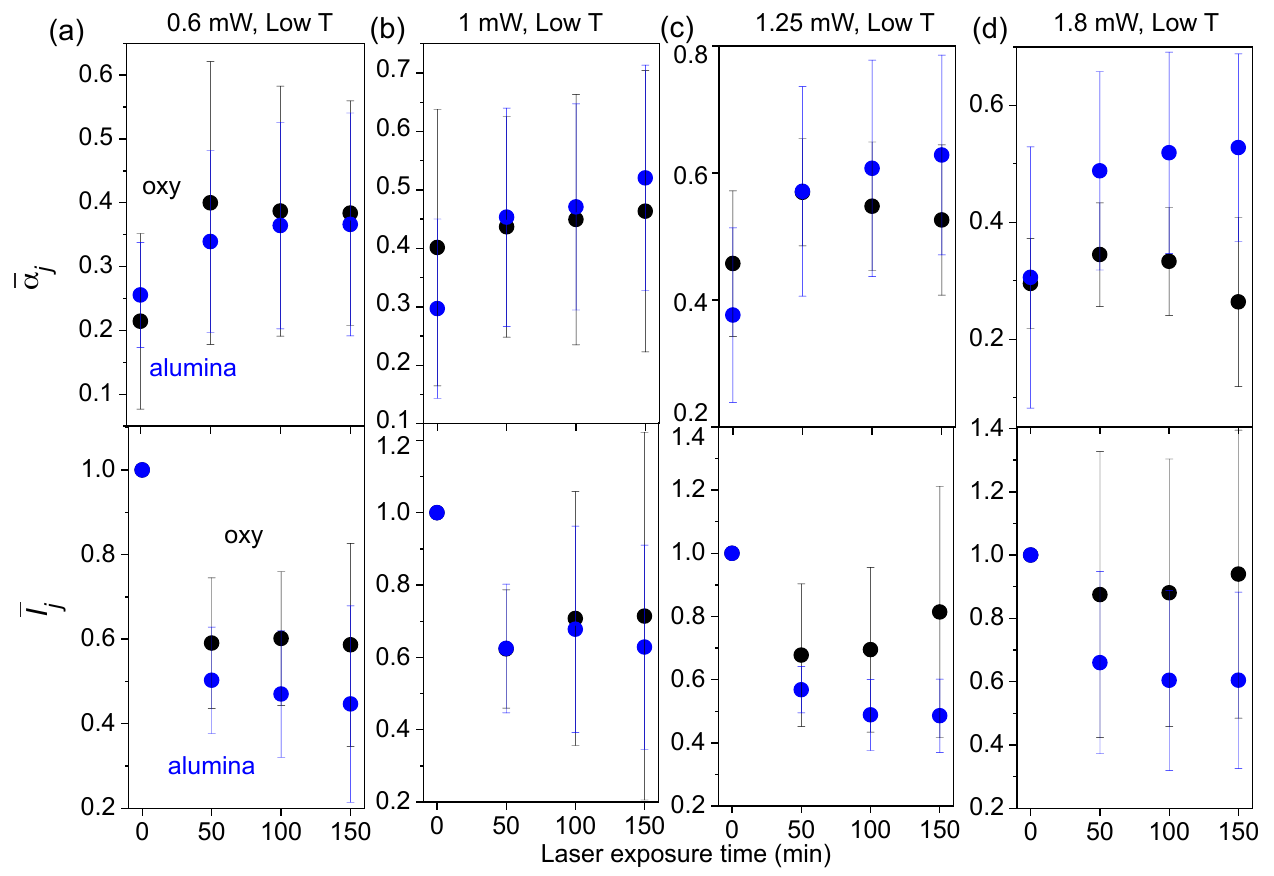}
    \caption{Top and bottom panels respectively show the evolution of average single photon purity ($\overline{\alpha _{j}}$) and average fluorescence intensity ($\overline{I _{j}}$) at laser powers of (a) 0.6~\milliwatt, (b) 1~\milliwatt{}, (c) 1.25~\milliwatt{}, and (d) 1.8~\milliwatt{} at low temperature.}
    \label{fig:S10}
\end{figure}
\begin{figure}[!t]
    \centering
    \includegraphics[width=0.75\linewidth]{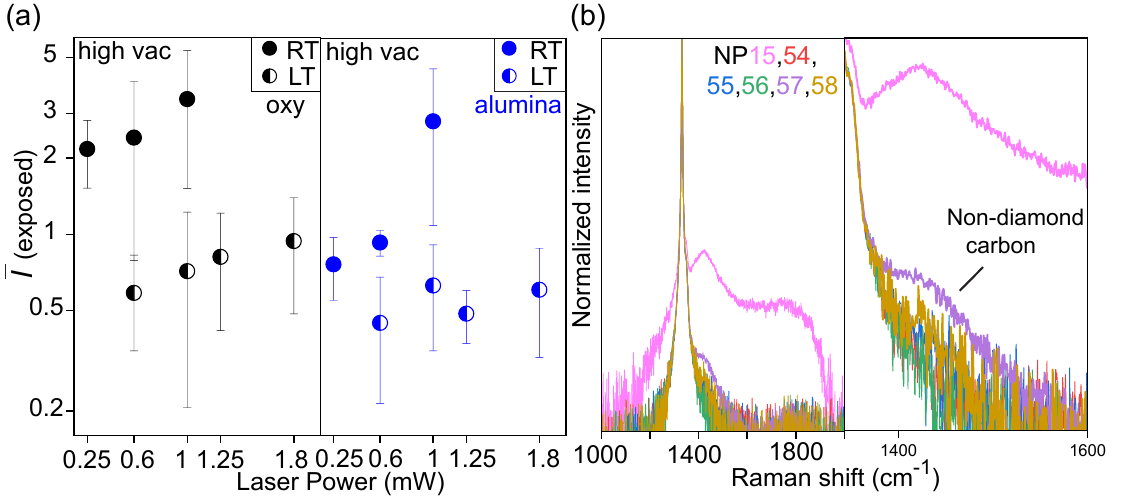}
    \caption{(a) Left and right panels show the evolution of average fluorescence intensity ($\overline{I}(\text{exposed})$) (plotted on a logarithmic scale) with laser power in high vacuum at different temperatures for oxygen- and alumina-terminated nanopillars, respectively. (b) Left panel shows the normalized Raman spectra of oxygen terminated nanopillars (NP54-58 were laser exposed at 1.8~\milliwatt{} at low T, while NP15 was laser exposed at 0.25~\milliwatt{} at room temperature under high vacuum). The signal intensity is plotted on a logarithmic scale to enhance the visibility of weak non-diamond features. The right panel shows an expanded view of the Raman spectra highlighting weak signatures of non-diamond carbon in NP57 and NP58.}
    \label{fig:S11}
\end{figure}
\FloatBarrier
\newpage
\section{Supplimentary References}
\bibliographystyle{aipnum4-1} 
\bibliography{references}     